\documentclass[prl,aps,superscriptaddress,twocolumn,floatfix,showpacs, showkeys]{revtex4-1}

\usepackage{graphicx}
\usepackage{epstopdf}
\usepackage{amsmath}

\usepackage{color}

\usepackage[utf8]{inputenc}

\begin{document}
\title{Giant Nernst effect in the incommensurate charge density wave state of P$_4$W$_{12}$O$_{44}$}

\author{Kamil K. Kolincio} 
\affiliation{Laboratoire CRISMAT, UMR 6508 du CNRS et de l'Ensicaen, 6 Bd Marechal
Juin, 14050 Caen, France.}
\affiliation{Faculty of Applied Physics and Mathematics, Gdansk University of Technology,
Narutowicza 11/12, 80-233 Gdansk, Poland}
\author{Ramzy Daou}
\affiliation{Laboratoire CRISMAT, UMR 6508 du CNRS et de l'Ensicaen, 6 Bd Marechal
Juin, 14050 Caen, France.}
\author{Olivier Pérez}
\affiliation{Laboratoire CRISMAT, UMR 6508 du CNRS et de l'Ensicaen, 6 Bd Marechal
Juin, 14050 Caen, France.}
\author{Laurent Guérin
\affiliation{Institut de Physique de Rennes, UMR CNRS 6251, Bât. 11 A, B, C, E – 10B
263 av. Général Leclerc, 35042 Rennes, France.}}
\author{Pierre Fertey}
\affiliation{Société civile Synchrotron SOLEIL, L'Orme des Merisiers, Saint-Aubin - BP 48, 91192 GIF-sur-YVETTE, France.}

\author{Alain Pautrat}
\affiliation{Laboratoire CRISMAT, UMR 6508 du CNRS et de l'Ensicaen, 6 Bd Marechal Juin, 14050 Caen, France.}

\begin{abstract}
We report the study of Nernst effect in quasi-low dimensional tungsten bronze P$_4$W$_{12}$O$_{44}$ showing a sequence of Peierls instabilities. We demonstrate that both condensation of the electronic carriers in the CDW state and the existence of high-mobility electrons and holes originating from the small pockets remaining in the incompletely nested Fermi surface give rise to a Nernst effect of a magnitude similar to that observed in heavy fermion compounds.
\end{abstract}
\maketitle

When a longitudinal thermal gradient $\nabla T_x$ is applied to a solid in a presence of perpendicular magnetic field, it leads to the generation of a transverse electric field $E_y$. This is recognized as the Nernst effect. Until the discovery of non-zero Nernst coefficient $N=\frac{E_y}{\nabla T_x}$ in the pseudogap phase of high-T$_c$ cuprate superconductors (SC) \cite{Xu2000}, this phenomenon mostly attracted the attention of researchers exploring the motion of the vortices in the mixed state\cite{Huebener1995}. More recently, much effort has been devoted to elucidate the key parameters which control the magnitude of the Nernst effect. From general arguments based on a semiclassical model of electronic transport, its magnitude is expected to grow as the ratio of electronic mobility over Fermi energy, $\frac{\mu}{E_F}$ \cite{Behnia2016}. One can then distinguish the two factors favorable for the increase of $N$: small Fermi energy, or large electronic mobility. Both of these factors: large $\mu$ and reduced $E_F$ have been found to give a rise to giant features in the Nernst effect in peculiar heavy fermion compounds. The key examples are URu$_2$Si$_2$ \cite{Bel2004U}, CeCoIn$_5$ \cite{Bel2004Ce} and PrFe$_4$P$_{12}$ \cite{Pourret2006}. Recently, strong $N$ has also been observed in a number of low dimensional materials as organic Bechgaard salts \cite{Wu2003, Choi2005}, Li$_{0.9}$Mo$_6$O$_{17}$ \cite{Cohn2012} or graphene \cite{Checkelsky2009, Zuev2009}. In these cases, the $N$ values were above those expected within a simple diffusive picture of electronic transport.  

The charge density wave (CDW) state of a quasi-low dimensional solid shows, to some extent, similarities with dilute anisotropic metals. The Peierls instability is associated to the nesting of the Fermi surface (FS) accompanied by condensation of electronic carriers. While in an ideal 1D material the Fermi surface is expected to be completely destroyed, the nesting in 2D systems is not perfect and small FS pockets containing high mobility carriers can remain \cite{Gruner1988, Monceau2012}. Despite these strongly favorable conditions, the Nernst effect was explored only in a few of the CDW materials as NbSe$_2$\cite{Bel2003}, where the Peierls instability involves the nesting of the minor part of the Fermi Surface. In addition to that, in NbSe$_2$, the positive signal due to superconducting fluctuations dominates over the component driven by quasiparticles as soon as the temperature approaches $T_c$, similarly to the features found in cuprate superconductors \cite{Daou2009, DoironL2013}, where the relevance of CDW formation was recently emphasized \cite{Chang2012}.

The promising features standing behind the Peierls instability motivated us to explore the Nernst effect in a purely CDW material with strong FS nesting and absence of a superconducting transition. For this study, we have chosen P$_4$W$_{12}$O$_{44}$, which belongs to the monophosphate bronze family of quasi-2D metals known to exhibit Peierls instabilities.  In P$_4$W$_{12}$O$_{44}$, a sequence of three CDW transitions lead to the nesting of the major part of the high temperature Fermi surface  \cite{Canadell1991, Paul2014, Wang1989} with several pockets left at low temperatures. Those were found to contain light free electronic carriers with the mobility high enough to allow observation of Shubnikov de Haas (SdH) \cite{Beierlein2003} and de Haas-van Alpen \cite{LeTouze1995} oscillations in this material. This predisposes P$_4$W$_{12}$O$_{44}$ to exhibit large Nernst coefficient.

\begin{figure}
 \resizebox{0.9\columnwidth}{!}{\includegraphics{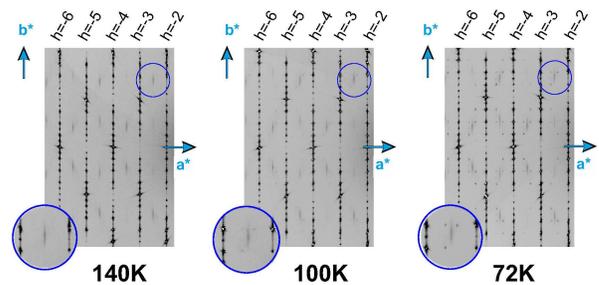}}%
 \caption{ \label{trans-1} Region of the (h0l)$^{\star}$ plane assembled from frames collected at $T$ = 140 K, $T$ = 100 K, $T$ = 72 K. The diffuse scattering observed for the position $\frac{\mathbf{a}^{\star}}{2}$ weakened noticeably above $T_{P1}$ \ while satellite reflections appear in irrational position. }
 \end{figure}

\begin{figure}
 \resizebox{0.9\columnwidth}{!}{\includegraphics{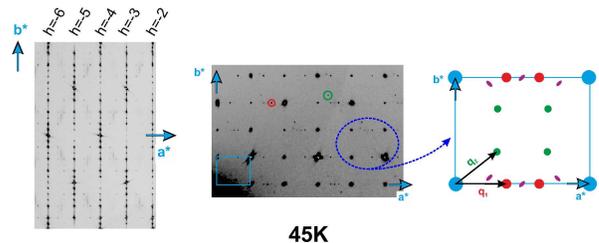}}%
 \caption{ \label{trans-2} Region of the (h0l)$^{\star}$ (left part) and (hk0)$^{\star}$ (middle part) planes assembled from frames collected at $T$=45K. The red and green circles show respectively satellite reflections associated to $\mathbf{q_1}$ and $\mathbf{q_2}$; the light blue rectangle correspond to the average unit cell. The right part is a schematic representation of the area blue-dashed circled in the (hk0)$^{\star}$ plane. Burgundy ellipsoids, red and green circles summarized all the diffraction features visible on the (hk0)$^{\star}$ plane.  }
 \end{figure}

To characterize the CDW transition, we have performed a high resolution thermal X-ray diffuse scattering experiment on a single crystal of P$_4$W$_{12}$O$_{44}$ showing an adequate crystalline quality (plate-like single crystals with typical size of 5 mm x 1 mm x 0.5 mm were grown by chemical vapor transport \cite{Teweldemedhin1991}). 
The experiment was performed at beamline Crystal of synchrotron Soleil, using monochromated
radiation with wavelength $\lambda$ = 0.50718 \AA~ and beam size
of 200 $\mu$m x 200 $\mu$m. Sample was cooled in a stream of cold He gas. Frames were collected above $T_{P1}$ = 120 K (temperature expected for the first transition), in between $T_{P1}$ \ and $T_{P2}$ = 62 K (temperature expected for the second transition), and below $T_{P2}$; however it was not possible to decrease the temperature of the sample below $T$ = 36 K and perform a diffraction experiment at temperatures lower than $T_{P3}\simeq 30$ K. Data were then gathered at $T$ = 293 K, 200 K, 140 K,  100 K, 72 K and  45 K respectively. The different sets of images were treated using
Crysalis~\cite{crysalis}. For the different temperatures, orientation matrix and cell parameters were calculated and oriented planes were assembled from the experimental frames. At $T$ = 293 K, above $T_{P1}$, P$_4$W$_{12}$O$_{44}$ \/ exhibits an orthorhombic symmetry (space group $P2_12_12_1$) with lattice parameters:
a = 5.3032(8) \AA, b = 6.5783(11) \AA \/ and c = 23.603(11) \AA. Above $T_{P1}$ \ diffuse segments running along $\mathbf{a}^{\star}$ are observed for the position $\frac{\mathbf{a^{\star}}}{2}$ (figure~\ref{trans-1}, $T$ = 140 K). At $T$ = 100 K satellite reflections appear from both sides of the diffuse segments; the intensity of these reflections is enhanced when the temperature decreases (figure~\ref{trans-1}, $T$ = 100 K and $T$ = 72 K). At $T$ = 45 K, below $T_{P2}$, (h0l)$^{\star}$ plane  remains unchanged (see left part of figure~\ref{trans-2}) but additional diffraction spots are observed on the (hk0)$^{\star}$ plane (see middle part of figure~\ref{trans-2}); these spots are drawn using green circles and burgundy ellipsoids in the scheme on the right part of the figure~\ref{trans-2}.

\begin{figure}
 \resizebox{1\columnwidth}{!}{\includegraphics{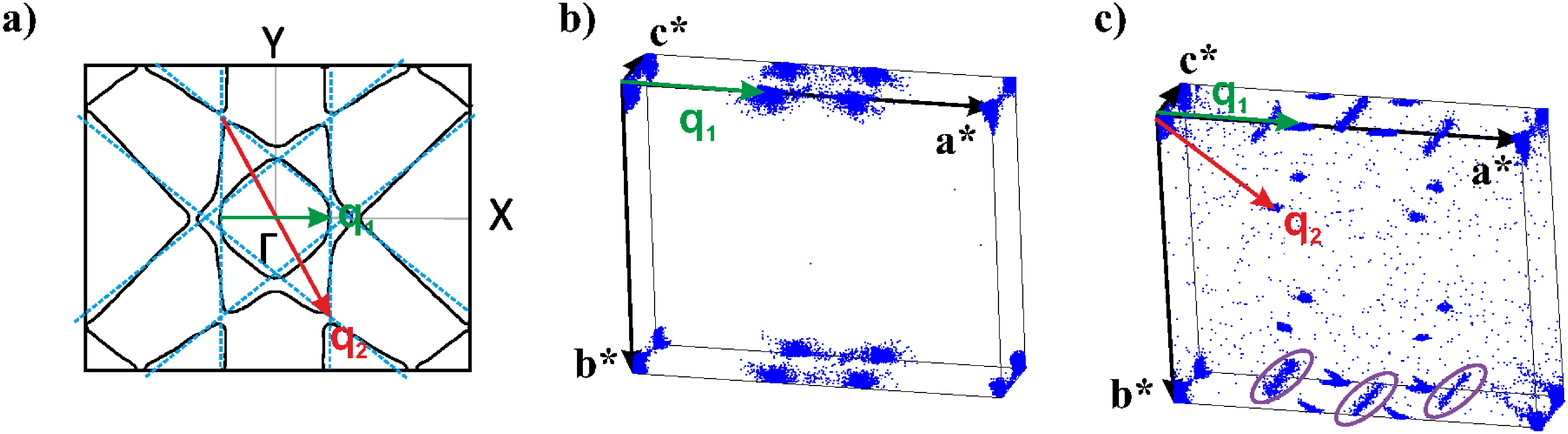}}%
 \caption{ \label{sum-1} a) Combined Fermi Surface (FS) of P$_4$W$_{12}$O$_{44}$ \ from \cite{Canadell1991}; the dashed blue lines show its decomposition into 3 FS and the $\mathbf{q_1}$ \ and $\mathbf{q_2}$ \ nesting vectors are reported. b) Projection into the origin reciprocal cell  of all the peaks extracted from the experimental X-ray diffraction frames collected at $T$ = 72 K. The cloud of peaks located on the $\mathbf{a^{\star}}$ \ edge of the cell requires the $\mathbf{q_1}$ \ vector to be indexed. c) Projection into the origin reciprocal cell of all the peaks extracted from the experimental X-ray diffraction frames collected at $T$ = 45 K. The clouds of peaks associated to $\mathbf{q_1}$  and $\mathbf{q_2}$ are clearly identified. The segments, purple circled, are diffuse scattering and correspond to the diffraction features drawn with burgundy ellipsoids in the right part of the figure~\ref{trans-2}. }
 \end{figure}

The positions of diffraction peaks on the experimental diffraction frames collected at $T$ = 72 K and 45 K were extracted using Crysalis and introduced into Jana2006~\cite{jana2006}. A subroutine of this program, allowing a manual cell determination, is useful to extract the average periodicity and superperiodicities. Once the average cell is defined from the intense reflections, all the diffraction peaks are projected in the reciprocal origin cell. For a regular crystal clouds of peaks are expected only at the node of the cell.
At $T$ = 72 K, additional clouds of reflections are observed along the $\mathbf{a^{\star}}$ edges of the cell (see figure~\ref{sum-1} b); the indexation of these reflections is requiring the introduction of a wave vector determined and refined with the Jana2006 subroutine as follow: $\mathbf{q_1}=0.3825(2) \mathbf{a^{\star}}$. The irrational value of the component of $\mathbf{q_1}$ reveals an one dimensional incommensurate modulated structure in this state.

At $T$ = 45 K, the diffraction phenomenon becomes mores complex; additionally to the peaks associated to $\mathbf{q_1}$, the (hk0)$^{\star}$ plane (figure~\ref{trans-2}) exhibits new diffraction features. The application of the Jana2006 procedure, described above, reveals their actual character; this is summarized 
in the figure~\ref{sum-1} c. Two types of diffuse scattering running along $\mathbf{c^{\star}}$ are observed: one is located at $\frac{\mathbf{a^{\star}}}{2}$ and the other one at 0.26 $\mathbf{a^{\star}}$ + 0.07 $\mathbf{b^{\star}}$. Supplementary peaks requiring the introduction of a second wave vector $\mathbf{q_2}=0.3099(4) \mathbf{a^{\star}} + 0.2908(3) \mathbf{b^{\star}} $ are also observed; they are characteristic of the second transition occurring below $T_{P2}$. The 
full diffraction pattern at $T$ = 45 K is compatible with a two dimensional incommensurate modulated structure where all the satellite reflections can be indexed using $\mathbf{q_1}$, $\mathbf{q_2}$ and $\mathbf{q_1} \pm \mathbf{q_2}$. $\mathbf{q_1}$ \ and $\mathbf{q_2}$ \ are in agreement with
wave vectors reported by Foury \textit{et al.} from X-ray diffuse scattering investigations\cite{Foury19931}; the diffuse line observed at $T$ = 45 K and located at 0.26 $\mathbf{a^{\star}}$ + 0.07 $\mathbf{b^{\star}}$ \ (figure~\ref{sum-1} c) could be correlated with the wave vector associated to the third transition expected below 30 K \cite{Canadell1991,Foury19931}.

\begin{figure}
 \resizebox{0.9\columnwidth}{!}{\includegraphics{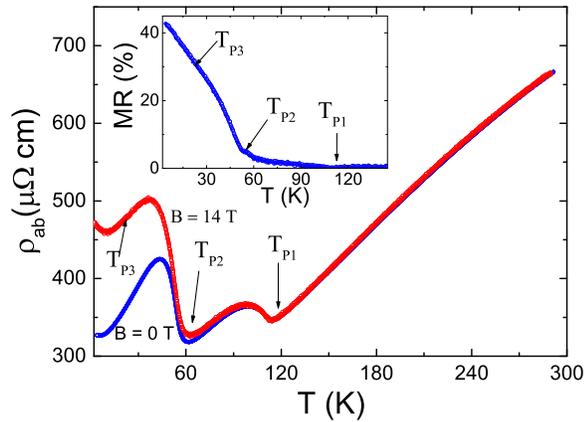}}%
 \caption{ \label{MR} Resistivity as a function of temperature measured with (red color) and without (blue color)  magnetic field applied perpendicularly to the (ab) plane. Inset: Magnetoresistance vs temperature measured at $B$ = 14 T.}
 \end{figure}

To investigate further the CDW transitions in P$_4$W$_{12}$O$_{44}$, we have studied the transport properties of this material. Figure \ref{MR} displays the in-plane resistivity measured both in absence, and presence of magnetic field using a standard four-probe method. The anomalies at $T_{P1}$ and $T_{P2}$ are pronounced in $\rho_{ab}$ as metal - metal transitions, which is typical  of the imperfect CDW nesting in a 2D metal. No anomaly is observed at $T_{P3}$ = 30 K. This temperature was previously identified from X-ray data \cite{Foury19931}, thermopower \cite{Hess1996} and Hall effect \cite{Rotger1994}. Above $T_{P1}$ the magnetoresistance ($MR$ = $\frac{\rho(B)-\rho(B=0)}{\rho(B=0)}$) is negligibly small. Upon cooling below $T_{P1}$, the presence of magnetic field increases the resistance by a few percent. Below $T_{P2}$, $MR$ increases significantly reaching even $\approx$ 40\% at low temperatures (see inset of fig. \ref{MR}). No significant anomaly is observed at $T_{P3}$. The gradual increase of magnetoresistance upon the transitions at $T_{P1}$ and $T_{P2}$ illustrates the process of subsequent reduction of the high temperature Fermi surface (fig. \ref{sum-1} a) to the small pockets containing high mobility carriers. The relevance of high mobility (several $10^3$cm$^2$V$^{-1}$s$^{-1}$) of both electrons and holes remaining in the unnested regions of the Fermi surface has been confirmed by the  Hall effect studies of this compound \cite{Rotger1994, Hess1996}.

\begin{figure}
 \resizebox{1\columnwidth}{!}{\includegraphics{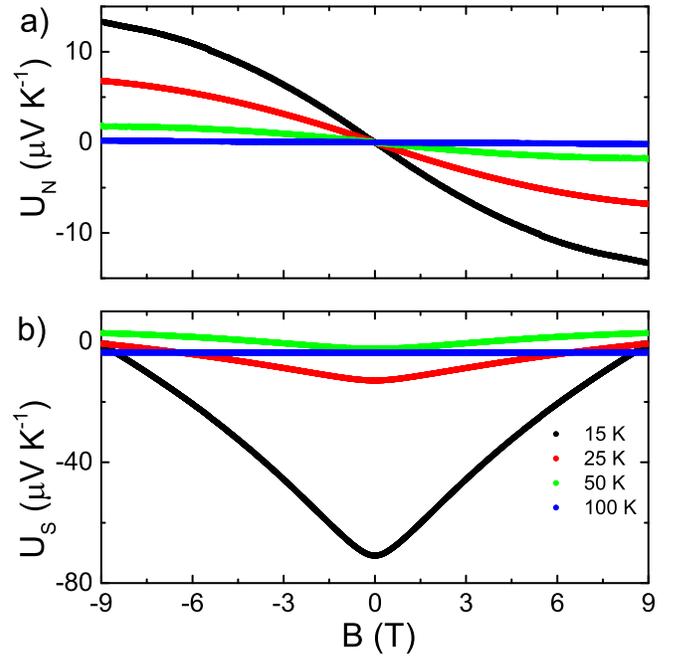}}%
 \caption{ \label{sweep} Field dependence of Nernst ($U_N$ - (a))and Seebeck ($U_S$ - (b)) signal components extracted from the experimental data. }
 \end{figure}

The Nernst effect measurements were conducted in the PPMS cryostat with a custom sample holder and electronics, using the standard one-heater two-thermometer technique. While the sample is glued to the substrate (SrTiO$_3$) with GE varnish to homogenize the thermal gradient, the temperature probes are attached directly to the sample. We are confident that the measured gradients are therefore indicative of those present in the sample, not the substrate. The use of an insulating, non-magnetic substrate to establish the thermal profile also ensures that the thermal gradient is independent of magnetic field. The magnetic field was applied parallel to \textit{c} and swept continuously from 9 T to -9 T. The Seebeck and Nernst voltage components, respectively symmetric and antisymmetric with the magnetic field have been separated. Typical results are shown in figure \ref{sweep}.

\begin{figure}
 \resizebox{1\columnwidth}{!}{\includegraphics{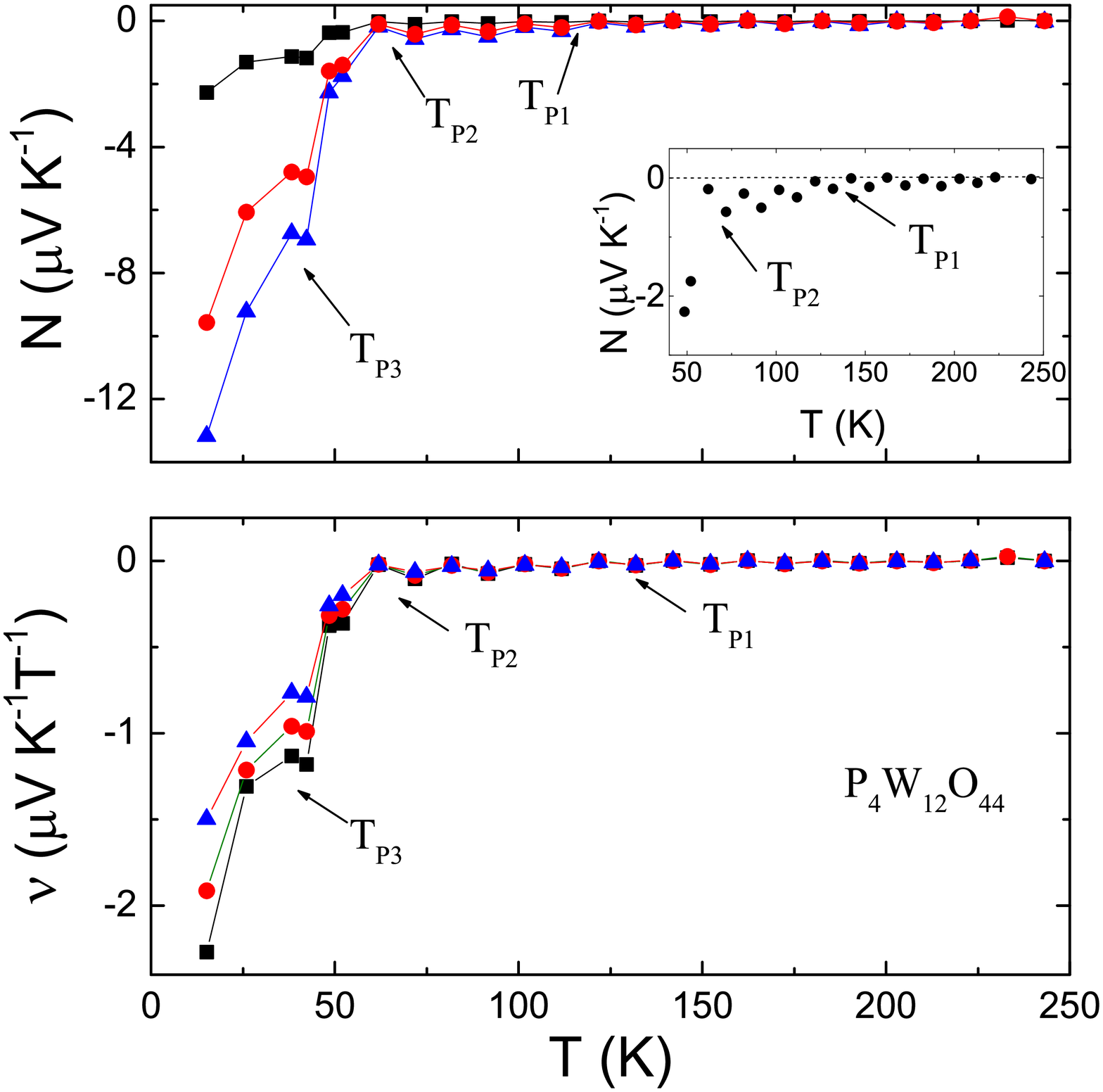}}%
 \caption{ \label{Nernstw} Nernst coefficient vs. temperature for P$_4$W$_{12}$O$_{44}$ measured in $B$ = 1 T (black squares), $B$ = 4 T (red circles), $B$ = 8.8 T (blue triangles), displayed in two scales: $N$  (upper panel) and $\nu=\frac{N}{B}$ (lower panel).  Inset: expanded view of the $N(T)$ dependence measured in $B$ = 8.8 T.}
 \end{figure}

The Nernst coefficient measured in the (ab) plane, plotted against temperature is shown in fig. \ref{Nernstw}. 
For $T >T_{P1}$, $N$  is roughly constant with $T$ and close to zero. Upon cooling below $T_{P2}$, $N$  becomes visibly negative (we use the vortex convention) and its magnitude increases as $T$ decreases. For $T <T_{P2}$ one can observe a significant downturn of $N(T)$, with a small minimum at  $T \approx$ 40 K, close to third CDW transition ($T_{P3}$ = 30 K) which is apparently easier to detect with Nernst effect than with resistivity. A contrast between the slight change of $N(T)$ for $T_{P2}< T < T_{P1}$ and a steep decrease of that value below $T_{P2}$ reflects well the anomalies observed in magnetoresistance and is in agreement with reported Hall data measured in the (ab) plane \cite{Rotger1994}, where the significant increase of negative Hall voltage below $T_{P2}$ was attributed to opening of the small electronic pockets in Fermi surface, containing high mobility carriers. Interestingly, a similar behavior was observed in thermopower with $\nabla$T applied in \textit{a} crystallographic direction \cite{Hess19974}. 

The relatively sharp change of $N(T)$ in P$_4$W$_{12}$O$_{44}$ at $T_{P2}$ (and to a lesser extent $T_{P1}$) contrasts with the behavior seen in another CDW material, quasi-2D NbSe$_2$ \cite{Bel2003}, where no sharp feature is detected in $N(T)$ at $T_{CDW}$. This is also true of the crossovers seen in the Nernst effect at the pseudogap line of the cuprate superconductors \cite{Daou2009}. CDW order seems to be a universal feature of the cuprate phase diagram but it is currently unclear whether it is the relevant order parameter underlying the pseudogap phase \cite{Chang2012}. In contrast to recalled examples, P$_4$W$_{12}$O$_{44}$ does not undergo a superconducting transition. Therefore we observe the Nernst signal of the pristine CDW state, not interrupted by superconducting fluctuations or involved in the SC-CDW competition. Comparing our case to NbSe$_{2}$, one should also emphasize the different impact of nesting properties in both materials. P$_4$W$_{12}$O$_{44}$ undergoes three Peierls transitions leading to almost complete destruction of the Fermi surface (\ref{sum-1} a) accompanied with the condensation of the majority of electronic carriers, while the CDW instability in NbSe$_2$ leads only to a marginal reconstruction of FS \cite{Rossnagel2001}. We also observe, that $N$ in P$_4$W$_{12}$O$_{44}$ is sublinear with $B$ over the whole temperature range below $T_{P1}$, which was also seen in the NbSe$_2$ in the CDW state. 

The value of $\nu \approx $ 2.4 $\mu$VK$^{-1}$T$^{-1}$ obtained by us at $T$ = 15 K and $B$ = 1 T parallels the scale of Nernst effect in heavy fermion compounds as URu$_2$Si$_2$ (maximum of normal quasiparticle signal of 4 $\mu$VK$^{-1}$T$^{-1}$ at 4 K)\cite{Bel2004U}, and even exceeds the magnitude of $\approx $ 1 $ \mu$VK$^{-1}$T$^{-1}$ reported in CeCoIn$_5$ at 3 K \cite{Bel2004Ce}. Our result is also close to the Nernst coefficient found in PrFe$_4$P$_{12}$ ($\approx$ 3 $\mu$VK$^{-1}$T$^{-1}$) at a similar temperature. At this point, we would like to mention that we have not observed any sign of saturation of $N(T)$. We suggest that it is reasonable to assume, that the magnitude of $N$ increases even further at lower temperatures. It was recently shown in URu$_2$Si$_2$ that the low-temperature maximum of N can have a strong sample dependence \cite{yamashita2015}. While in this case it has been attributed to exotic superconducting fluctuations, this highlights the difficulty in obtaining the correct value of the normal state Nernst signal in the low temperature limit. In our case there is no sign of superconductivity, but other effects such as phonon drag are not well understood and may cause a sample dependent signal at low temperatures \cite{Behnia2015}.

One approximate approach \cite{Behnia2009} predicts that the values of $N$ follow the scaling law:
\begin{equation}
\label{Nernst1}
\frac{N}{BT}=\frac{\pi^2 k_B \mu}{3eT_F}
\end{equation}
where $\mu$ is the electronic mobility and $T_F$ the Fermi temperature.
The value of $\frac{N}{BT}\approx$ 0.13 $\mu$VK$^{-2}$T$^{-1}$ at $T$ = 15 K was evaluated using $\mu$ from Ref. \onlinecite{Hess1996} for relevant temperature and $T_F$ estimated from the Fermi wavevector corresponding to a pocket revealed by SdH oscillations in this material\cite{LeTouze1995, Beierlein2003}. Interestingly, the predicted value stands in a very good agreement with the experimental result $\frac{N}{BT}=$ 0.15 $\mu$VK$^{-2}$T$^{-1}$. This shows that the magnitude of the Nernst effect in the low temperature state is well accounted for by considering the reduction of the carrier number. The Nernst effect can also be enhanced in well-compensated materials, as was seen in NbSe$_2$ \cite{Bel2003}, where the Nernst signal is maximised just as the Hall coefficient changes sign. P$_4$W$_{12}$O$_{44}$ has also been reported to be a compensated metal in the CDW state with FS containing both electron and hole pockets in the CDW state \cite{Hess1996, Rotger1994, Wang1989}. In a perfectly compensated material, the Nernst coefficient can be described with the following expression\cite{Oganesyan2004}: 
\begin{equation}
N= \frac{2\pi^2}{3} \frac{k_B^2T\tau}{e \hbar} \frac{1}{(k_Fl_B)^2}
\end{equation}
Inserting the magnetic length $l_B=\sqrt{\frac{\hbar}{eB}}$ and the scattering time $\tau=\frac{l_em^*}{k_F}$, one obtains: 
\begin{equation}
N= \frac{2\pi^2}{3} \frac{k_B^2TBl_em^*}{\hbar^3k_F^3}
\end{equation}
To estimate the mean free path $l_e$, we have fit the $T <$ 50 K part of $N(T)$ plot at $B$ = 1 T using values of $m^*=0.45$ times the free electron mass ($m_0$) and $k_F=0.64$ $nm^{-1}$ evaluated from Shubnikov - de Haas oscillations reported in this material \cite{LeTouze1995}. We have found the value of $l_e=265$ \AA~ which compares very well to the quantity of 260 \AA~ estimated from Dingle temperature. Since this model is valid only in the limit of low temperatures, and does not account for phonon effects, this surprisingly good agreement should not be overinterpreted. Nevertheless it is worth mentioning that both values of $l_e$ are an order of magnitude greater than the lattice constants $a$ and $b$. One shall also mention that, his large mean free path is sufficiently high to observe quantum oscillations in this compound.


In conclusion, we report on the study of Nernst effect in a quasi-2D metal showing charge density wave instabilities. This effect induces the partial destruction of the
Fermi surface, and the reduction of the carrier concentration. This results in the electrical compensation due to both electron and hole pockets left in the FS and the high mobility of the carriers originating from them, leading to the enhanced Nernst effect with a magnitude as large as those reported in heavy fermion compounds. We also suggest, that since several other members of the monophosphate bronzes family undergo the Peierls transition above 300 K \cite{Roussel2001}, the search of large Nernst effect 
at room temperature in these materials appears promising.

Financial support by the French National Research Agency
ODACE ANR-Project number 2011-BS04-004-01 is gratefully
acknowledged.
%
 \end{document}